\documentclass{aa}
\usepackage{graphicx,epsf}

\newcommand{\accrh}{\hbox{$a_{\rm X}$}}
\newcommand{\accro}{\hbox{$a_{\rm Y}$}}
\newcommand{\desh}{\hbox{$d_{\rm X}$}}
\newcommand{\khh}{\hbox{$\kappa_{\rm X,X}$}}
\newcommand{\koh}{\hbox{$\kappa_{\rm X,Y}$}}
\newcommand{\koo}{\hbox{$\kappa_{\rm Y,Y}$}}

\newcommand{\hyd}{\hbox{X$_2$}}
\newcommand{\oxy}{\hbox{Y$_2$}}
\newcommand{\nh}{\hbox{$\left<N(\mathrm{X})\right>$}}
\newcommand{\no}{\hbox{$\left<N(\mathrm{Y})\right>$}}

\newcommand{\khdash}{\hbox{$\kappa_{\mathrm{X}}'$}}
\newcommand{\kodash}{\hbox{$\kappa_{\mathrm{Y}}'$}}
\newcommand{\tsh}{\hbox{$t_{\rm s}({\rm X})$}}
\newcommand{\tso}{\hbox{$t_{\rm s}({\rm Y})$}}
\newcommand{\ratemolh}{\hbox{$\Gamma(\mathrm{X}_2)$}}
\newcommand{\rateoh}{\hbox{$\Gamma(\mathrm{XY})$}}
\newcommand{\ratemolo}{\hbox{$\Gamma(\mathrm{Y}_2)$}}

\begin{document}

\title{Diffusive grain-surface chemistry involving the atoms and
  diatomic molecules of two elements}
\titlerunning{Diffusive grain surface chemistry}

\author{J.G.L. Rae \inst{1} \and N.J.B. Green\inst{2} \and
  T.W. Hartquist\inst{1} \and M.J. Pilling\inst{3} \and T. Toniazzo\inst{1,4}}
\institute{Department of Physics and Astronomy, University of Leeds, Leeds 
LS2 9JT, UK \and Chemistry Department, Kings College London, London
  WC2R 2LS, UK \and School of Chemistry, University of Leeds, Leeds
  LS2 9JT, UK \and Met.\ Office, London Road, Bracknell, Berkshire,
  RG12 2SZ, UK}
\offprints{J.G.L. Rae \email{jglr@ast.leeds.ac.uk}}
     \date{Received <date> / Accepted <date>}

\abstract{
A model of the grain surface chemistry involving the accretion of
atoms of two different elements, X and Y, and their reactions to form
species X$_2$, XY, and Y$_2$ was examined for a wide range of choices
for the values of its three free parameters -- the accretion rate of X
and Y, the desorption rate of X and the grain surface sweeping time
of Y, all considered relative to the grain surface sweeping rate of X.
Relative production rates of the diatomics were calculated with five
methods involving, respectively, a high-order truncation of the master
equation, a low-order truncation of the master equation, the standard
deterministic rate equation approach, a modified rate equation
approach and a set of approximations which are in some cases
appropriate for accretion dominated chemistry.  
The accuracies of the relative production rates calculated with the
different methods were assessed for the wide range of model parameters.
The more accurate of the low-truncation master equation calculations
and the standard deterministic rate equation approach gives results
which are in most cases within ten or twenty per cent of the results
given by the high-truncation master equation calculations.
For many cases, the more accurate of the low order truncation and the
standard deterministic rate equation approaches is indicated by a
consideration of the average number of atoms of the two species on the
grain's surface.


}

\authorrunning{J.G.L. Rae et al.}

\maketitle

\section{Introduction}
\label{intro}

If species accrete from the gas phase onto the
surfaces of grains more quickly than they diffuse across a grain
surface and react with each other, the surface chemistry is said to be
taking place in the {\it reaction limit}.  Such chemistry has been
treated by Pickles \& Williams
(1977), who used a system of rate equations similar to those employed in
gas-phase chemistry.  This approach is often referred to as the 
{\it standard deterministic rate equation method}. 

In some cases, species diffuse
across grain surfaces and react with each other faster than they
accrete.  Therefore, when a species is accreted, and there is already
another reactive species on the grain, they are likely to react before
another particle accretes.  The rates of reaction are therefore
limited by the rates of accretion;  the chemistry is said to be
occurring in the {\it accretion limit}.  Under these conditions, the
average numbers of reactive species on a grain are small, and a
stochastic model is required.  The first attempts at stochastic
modelling of interstellar grain surface chemistry
in the accretion limit were made by Allen \& Robinson (1977) and
Tielens \& Hagen (1982).  The model of Allen \& Robinson
(1977) was constructed under the assumption that when a molecule is 
produced in a grain surface reaction, it immediately enters the gas
phase;  the model was extended by Tielens \& Hagen
(1982) to allow for the accumulation of grain mantles.  In their model,
Tielens \& Hagen (1982) used  Monte Carlo techniques to calculate the
steady-state concentrations of mantled species.  Later, various other
authors constructed Monte Carlo models of grain surface chemistry
(e.g.\ Tielens \& Allamandola 1987; Tielens 1995 (unpublished);
Charnley et al.\ 1997; Tielens \& Charnley 1997; Charney 2001).
Charnley (1998) used Monte Carlo methods to solve the master equation
(which gives the probability that there is a certain number of
particles of each species) governing gas-phase chemistry, and
suggested that a similar approach may be applicable to grain surface
chemistry.

As Monte Carlo simulations are computationally too expensive to use in
the study of large systems, Caselli et al.\ (1998; 2002) and Shalabiea
et al.\ (1998) introduced semi-empirical modifications to the standard
determininstic rate equations, with the aim of solving the problem of
inaccuracy of rate equation methods in the accretion limit.

Biham et al.\ (2001) and Green et al.\ (2001) studied, with a master
equation approach, the production rate of H$_2$ on surfaces.  Green et
al.\ (2001) used a generating function method to obtain an analytic
solution for the H$_2$ production rate.  They also
investigated more complex chemistries, using sparse matrix
techniques to obtain the probabilities that a grain contains
given numbers of particles of each species.   With this approach, the
calculation must be truncated at a certain maximum number of particles
of each species, i.e.\ there is some {\it truncation value} for the
number of particles, above which the probabilities are assumed to be
zero. With a high enough truncation value, the results are almost
exact. Green et al.\ (2001) used the probabilities obtained in this
way to calculate the rates of production of molecules on grain
surfaces.  They solved the master equation numerically for the system
of H and O, reacting on grain surfaces to form H$_2$, O$_2$, and OH,
and also for the system of H, O, and N, reacting to form H$_2$, O$_2$,
OH, NH, N$_2$, NO, NO$_2$, H$_2$O, and NH$_3$.  They compared the
results given by the solution of the master equation
with a high truncation value, to those given by (i) the use of
the master equation method with a low truncation value; (ii) the standard
deterministic rate equation; and (iii) an approximate method, based on
the assumption that the sweeping rate of atomic hydrogen on grain
surfaces is so great that 
a reactive species already on the surface of a grain will react as
soon as an H atom is accreted.

Stantcheva et al.\ (2002) solved the master equation for the system of
H, O, and CO reacting on the surfaces of grains to produce O$_2$,
H$_2$, H$_2$O, CO$_2$, H$_2$CO, and CH$_3$OH.  They investigated the
accuracy of results given by the method for different truncation
values and also the accuracies of results given by other approaches.

In this paper, we examine the system of atoms X and Y reacting to form
X$_2$, XY, and Y$_2$ on grain surfaces.  We study a wide variety of
adsorption, desorption, and diffusion rates, covering the accretion
limit, the reaction limit, and cases intermediate between them. We 
compare the almost exact results, obtained through the solution of
the master equation for high truncation values, with the results
obtained (i) through the solution of the master equation with low
truncation values;
(ii) with the approximate method used by Green et al.\ (2001); (iii) 
with the standard deterministic rate equation approach; and (iv) with a 
modified rate equation approach similar to that employed by Caselli et 
al.\ (2002). Although the network we study is smaller than that of 
Stantcheva et al.\ (2002), we examine a much larger range of
adsorption, desorption, and diffusion rates, thereby gaining
insight into the parameter ranges in which each method may be applied.

In Sect.\ \ref{appr}, the various approaches to the problem are discussed.
We present our results in Sect.\ \ref{results}, 
and Sect.\ \ref{conc} concludes the paper.

\section{The various approaches to the problem}
\label{appr}
We consider a system in which atoms X and Y accrete from the gas
phase onto the surfaces of grains, diffuse across the grain surfaces,
and react with each other to form X$_2$, Y$_2$, and XY, i.e.\ in which
the following reactions take place:
\begin{eqnarray}
\label{h2prod}
\rm X + X & \rightarrow & \hyd \\
\label{ohprod}
\rm Y + X & \rightarrow & \rm XY \\
\label{o2prod}
\rm Y + Y & \rightarrow & \oxy. 
\end{eqnarray}
We assume that the X$_2$, XY, and Y$_2$ molecules
remain on the grain surfaces, but that they do not
react with each other or with X or Y.  We allow for desorption of X
from grains, but assume that desorption of Y is negligible.
Let \accrh\ and \accro\ be the rates of accretion of X and Y
respectively, and
\desh\ be the rate of desorption of X.  Let \tsh\ and \tso\ be the grain
surface sweeping times for X and Y
respectively.  
To reduce the number of free parameters, we assume
\begin{equation}
\label{aheqao}
\accrh=\accro 
\end{equation}
This assumption is likely to be reasonable at some densities in dense
cores if X is hydrogen and Y is oxygen.  Furthermore, in a number of
previous studies (e.g.\ Stantcheva et al.\ 2002), the ratio of
accretion rates has been varied.

We normalise with respect to \tsh\ as follows.
\begin{eqnarray}
\label{alphaeq}
\alpha & = & \accrh\tsh \\
\label{deltaeq}
\delta & = & \desh\tsh \\
\label{taueq}
\tau & = & \frac{\tso}{\tsh} 
\end{eqnarray}
The dimensionless rates for Reactions (\ref{h2prod}) to (\ref{o2prod}) are
\begin{eqnarray}
\label{khh2}
\kappa_{\mathrm{XX}} & = & 2 \\
\label{kohtau}
\kappa_{\mathrm{XY}} & = & 1 + \tau^{-1} \\
\label{kootau}
\kappa_{\mathrm{YY}} & = & 2\tau^{-1}
\end{eqnarray}
Eqs.\ (\ref{khh2}) to (\ref{kootau}) may be compared to the rates
given by Eq.\ (9) of Caselli et al.\ (1998) for two arbitrary species
I and J.

The production rates of X$_2$, XY, and Y$_2$ were calculated with the
various approaches for $13\times 13\times 13$ sets of values of
$\alpha$,$\delta$,$\tau$  between 0.1 and 10.  Calculations were also
performed for $9\times 13\times 9$ sets of $0.01\leq\alpha\leq
0.09$, $0.1\leq\delta\leq 10.0$, $0.01\leq\tau\leq 0.09$.

We now discuss in detail the master equation approach, the standard
and modified rate equation approaches, and the approximate method of
Green et al.\ (2001). 

\subsection{Master equation approach}
\label{me}

Green et al.\ (2001) gave the master equation for the hydrogen and
oxygen system in their Eq.\ (29), and the rates of production of
H$_2$, OH, and O$_2$ in their Eqs.\ (26) to (28). The generalisation
of their equations to species X, Y, X$_2$, XY, and Y$_2$ is trivial.
The average numbers of X and Y atoms on the surface of a
grain are:
\begin{eqnarray}
\label{nheqn}
\nh & = &  \sum_{\stackrel{i=0}{j=0}}^{\infty}iP(i,j) \\
\label{noeqn}
\no & = &  \sum_{\stackrel{i=0}{j=0}}^{\infty}jP(i,j) 
\end{eqnarray}
where $P(i,j)$ is the probability that on the surface of a grain there
are $i$ atoms of species X and $j$ atoms of species Y.
We assumed the system to be in steady state, so that  
$\frac{{\rm d} P(i,j)}{\rm{d}t}=0$ for all $i,j$, and followed Green
et al.\ (2001) in solving the master equation through the inversion of
a sparse matrix by the row-indexed method.  As discussed in Sect.\
\ref{intro}, with this approach it is necessesary to
truncate the sums at particular values of $i$ and $j$.  They were
truncated first at $i=j=2$ (which will be referred to as the {\it low 
truncation case}), and then at higher values (which will be referred to as
the {\it high truncation case}), above which the results did not
change appreciably if $i$ and $j$ were increased further.  The results
given by the master equation approach with high truncation values of
$i$ and $j$ -- typically $i=j\ge 5$ --
can be considered to be exact, and will be referred to as
the {\it exact results}.

\subsection{Standard rate equation approach}
\label{re}
The standard deterministic rate equations for the H, O, H$_2$, OH, and
O$_2$ system are given by Caselli et al.\ (1998) in their Eqs.\ (4) to
(8).  They can be generalised easily to the X, Y, X$_2$, XY, and Y$_2$ system.
The rate equations for X and Y were integrated with a Gear
algorithm until the system
reached steady state.  Then, the production rates of X$_2$, XY, and
Y$_2$ were calculated from their rate equations.

\subsection{Modified rate equation approach}
\label{mre}
We introduced modifications to the rate equations, similar to
those suggested by Caselli et al.\ (1998, 2002) in their attempt to
develop a set of deterministic equations appropriate in both the reaction
limit and the accretion limit.  The modifications we used were based
on those used by Caselli et al.\ (2002).
%

We define $\theta_{1}$ and $\theta_{2}$ as
\begin{eqnarray}
\label{theta1}
\theta_{1} & = & \frac{\khh\nh+\koh\no}{\alpha} \\
\label{theta2}
\theta_{2} & = & \frac{\koo\no+\koh\nh}{\alpha}
\end{eqnarray}
In cases in which $\theta_{1}\leq 1$ and $\theta_{2}\leq 1$, the
standard deterministic rate equations were used.  Otherwise, the
equations were modified as follows.

In the case that $\theta_{1}> 1$ and $\theta_{2}\leq 1$, the
probability that species Y reacts is either less than or equal to 1,
so that for Y$_2$ formation the standard deterministic rate equation
approach
can be used.  However the probability that species X reacts is greater
than 1, and the equations for the production of X$_2$ and XY must
be modified.  The rate coefficients \khh\ and \koh\ are replaced by
the larger of $\alpha$ and $\delta$ (Caselli et al.\ 1998, 2002);  this
quantity will be denoted by \khdash.  In addition, the formation rate
of X$_2$ is multiplied by the probability that X reacts with another X
instead of with a Y, and the rate of formation of XY is multiplied by
the probability  that X reacts with Y and not with another X
(following Caselli et al.\ 2002).  The production rates,
$\Gamma$(X$_2$) and $\Gamma$(XY), of X$_2$ and
XY, in this approach become
\begin{eqnarray}
\label{mre2_h2}
\Gamma(\mathrm{X}_2) & = & \khdash \nh \frac{\khh\nh}{\khh\nh+\koh\no}\\
\label{mre2_oh}
\Gamma(\mathrm{XY}) & = & \khdash\nh\frac{\koh\no}{\koh\no+\khh\nh}
\end{eqnarray}
After (\ref{mre2_h2}), (\ref{mre2_oh}), and the standard rate equation
for Y$_2$ are used, $\theta_{2}$ is recalculated; if it is found to
exceed unity then the approach described below for cases in which
$\theta_{1}>1$ and $\theta_{2}>1$ is applied.

The case in which $\theta_{1}\leq 1$ and $\theta_{2}> 1$ is the same
as the previous case, except that X is replaced by Y and vice versa.
So $\Gamma$(X$_2$) can be calculated with the standard determistic rate
equation approach, and the rates of formation, $\Gamma$(XY) and
$\Gamma$(Y$_2$), of XY and Y$_2$, are
\begin{eqnarray}
\label{mre3_oh}
\Gamma(\mathrm{XY}) & =
&\kodash\no\frac{\koh\nh}{\koh\nh+\koo\no}\\
\label{mre3_o2}
\Gamma(\mathrm{Y}_2) & = & \kodash\no\frac{\koo\no}{\koo\no+\koh\nh}
\end{eqnarray}
where \kodash is the larger of the accretion rate and the desorption
rate of Y;  however because we are neglecting desorption of Y,
$\kodash=\alpha$.  After Eqs.\ (\ref{mre3_oh}) and (\ref{mre3_o2}) and
the standard rate equation for X$_2$ production are used,
$\theta_{1}$ is recalculated; if it is found to be greater than
unity then the approach described below for cases in which
$\theta_{1}>1$ and $\theta_{2}>1$ is applied.

In the case in which both $\theta_{1}$ and $\theta_{2}$ are greater than
1, the rate equations for X$_2$, XY, and Y$_2$ must be modified.
For the formation rates of X$_2$ and Y$_2$ we used Eqs.\
(\ref{mre2_h2}) and (\ref{mre3_o2}), respectively.  For the XY
formation rate, we used Eq.\ (\ref{mre2_oh}) when $\theta_{1} >
\theta_{2}$, and Eq.\ (\ref{mre3_oh}) when $\theta_{1} < \theta_{2}$.

In all cases, the rates of change of \nh\ and \no\ are
\begin{eqnarray}
\label{mre_h}
\frac{\mathrm{d}\nh}{\mathrm{d}t} & = & \alpha - \delta\nh -
2\Gamma(\mathrm{X_2}) - \Gamma(\mathrm{XY}) \\
\label{mre_o}
\frac{\mathrm{d}\no}{\mathrm{d}t} & = &  \alpha -
2\Gamma(\mathrm{Y_2}) - \Gamma(\mathrm{XY})
\end{eqnarray}

Eqs.\ (\ref{mre_h}) and (\ref{mre_o}) were integrated with a Gear
algorithm until the system reached steady state.  At each
step in the integration, the rates of production of X$_2$, XY,
and Y$_2$ were calculated using the standard deterministic, or
modified, rate equations as appropriate, and were used in the
calculations of $\frac{\mathrm{d}\nh}{\mathrm{d}t}$ and
$\frac{\mathrm{d}\no}{\mathrm{d}t}$.

\subsection{Approximate Method}
\label{am}

Green et al.\ (2001) introduced an analytic approximation to the
master equation method, based on the assumption that the sweeping rate
of X is so fast that if there is a reactive species (X or Y in this
case) on the surface of a grain, a X atom accreting onto the grain
will react immediately.  This is equivalent to assuming that
$\khh\gg\alpha$, and that the probability of there
being more than one reactive species on the surface of a grain is
small.  Green et al.\ (2001) showed that, under
these assumptions
\begin{eqnarray}
\label{amh2}
\Gamma(\mathrm{X}_2) & \approx & \frac{\alpha^2
  P(0,0)}{2\alpha+\delta} \\
\label{amoh}
\Gamma(\mathrm{XY})
& \approx & \alpha^{2} \left[ \frac{1}{2\alpha+\delta} + \frac{1}{2\alpha}\right] P (0,0) \\
\label{amo2}
\Gamma(\mathrm{Y}_2)
& \approx & \frac{\koo}{\koo+2\alpha} \frac{\alpha}{2} P (0,0) 
\end{eqnarray}
where
\begin{equation}
P (0,0) \approx \left[ \frac{3}{2} + \frac{\alpha}{2\alpha + \delta}  \right]^{-1}
\end{equation}

\section{Results and discussion}
\label{results}
For each of the methods described in Sect.\ \ref{appr}, the production
rates of X$_2$, XY, and Y$_2$ were evaluated with $\alpha$,
$\delta$, \khh, \koh, and \koo\ given by Eqs.\ (\ref{alphaeq}) to
(\ref{kootau}), with $\alpha$, $\delta$, and $\tau$ having various
values, as described in Sect. \ref{appr}.  

\begin{figure}[!h]
\includegraphics[angle=0,width=7cm]{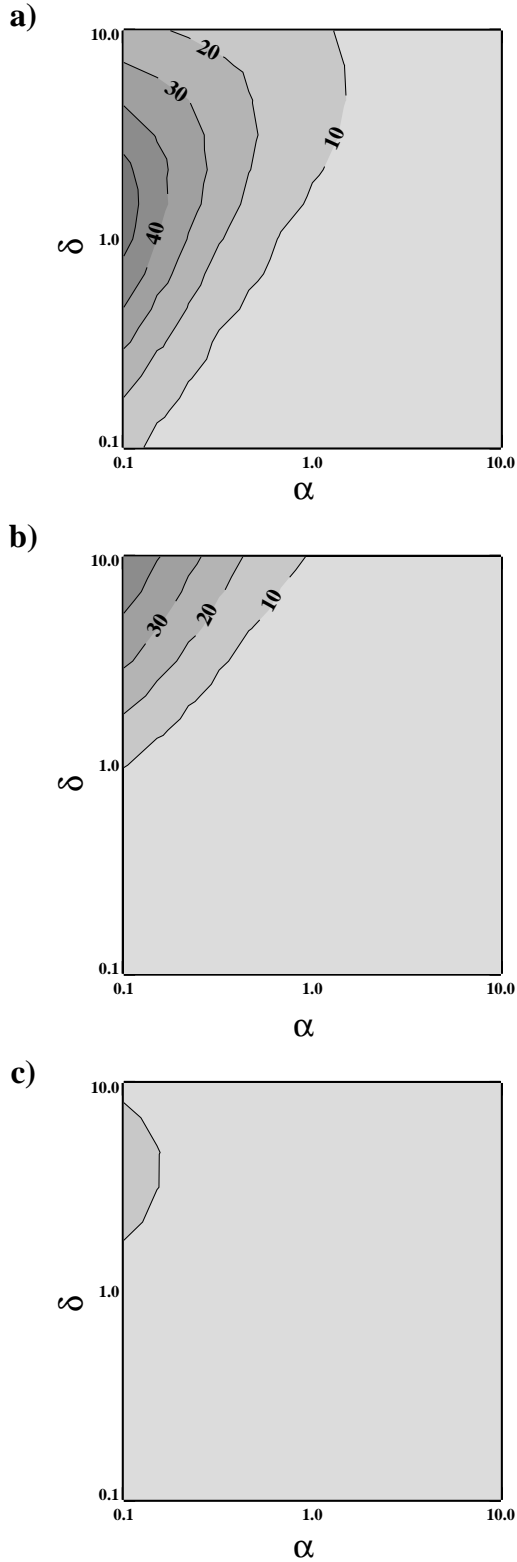}
\caption{Percentage discrepancies in results given by standard rate
  equation approach for $\tau=1.0$. a: discrepancy in $x$(X$_2$); b:
  discrepancy in $x$(XY); c: discrepancy in $x$(Y$_2$).
Darker regions indicate greater discrepancies.}
\label{refig}
\end{figure}

\begin{figure}[!h]
\includegraphics[angle=0,width=7cm]{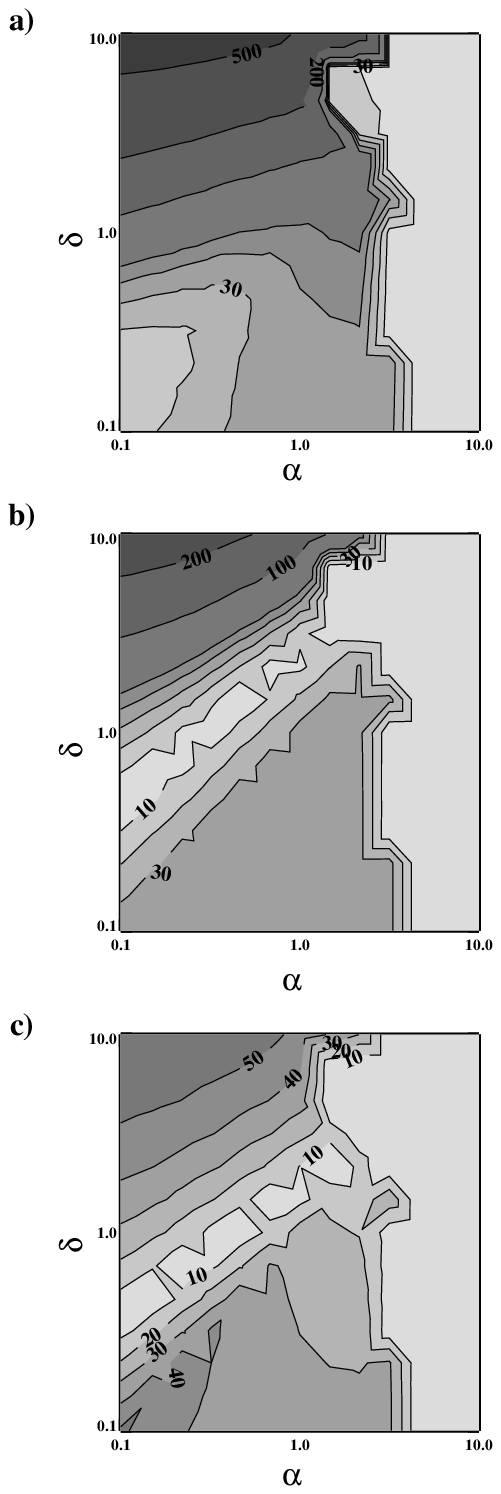}
\caption{Percentage discrepancies in results given by modified rate
  equation approach for $\tau=1.0$. a: discrepancy in $x$(X$_2$); b:
  discrepancy in $x$(XY); c: discrepancy in $x$(Y$_2$).
  Darker regions indicate greater discrepancies.}
\label{mrefig}
\end{figure}

\begin{figure}[!h]
\includegraphics[angle=0,width=7cm]{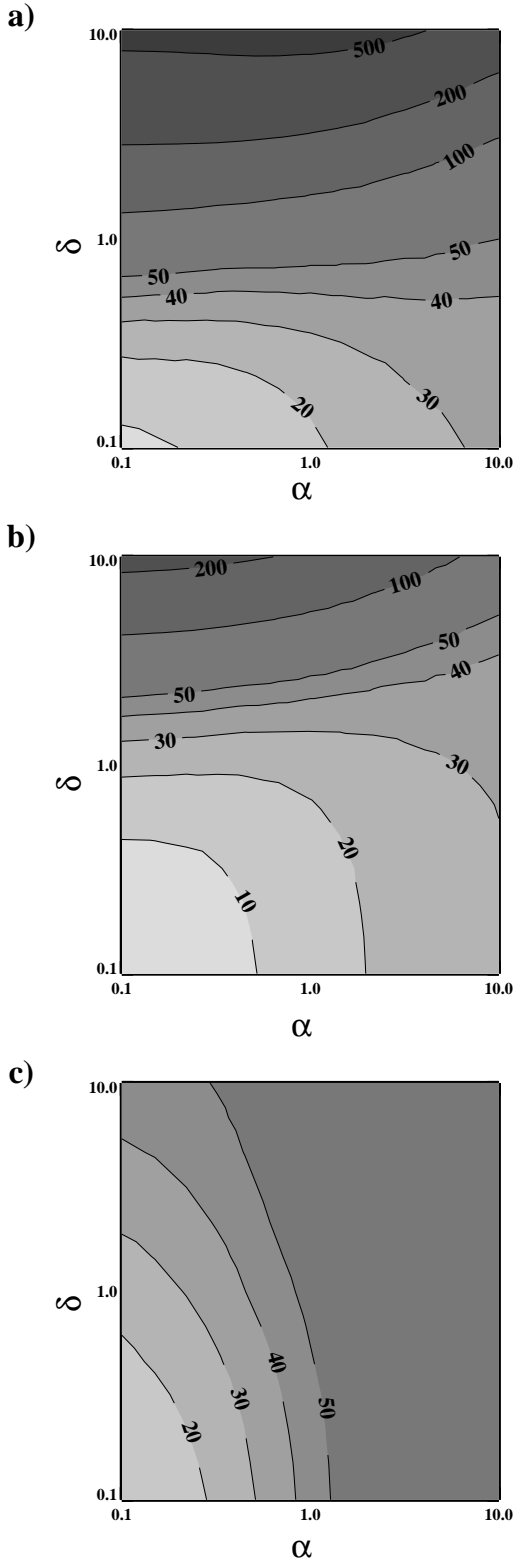}
\caption{Percentage discrepancies in results given by approximate 
 method for $\tau=1.0$. a: discrepancy in $x$(X$_2$); b:
  discrepancy in $x$(XY); c: discrepancy in $x$(Y$_2$).  
 Darker regions indicate greater discrepancies.}
\label{amfig}
\end{figure}

\begin{figure}[!h]
\includegraphics[angle=0,width=7cm]{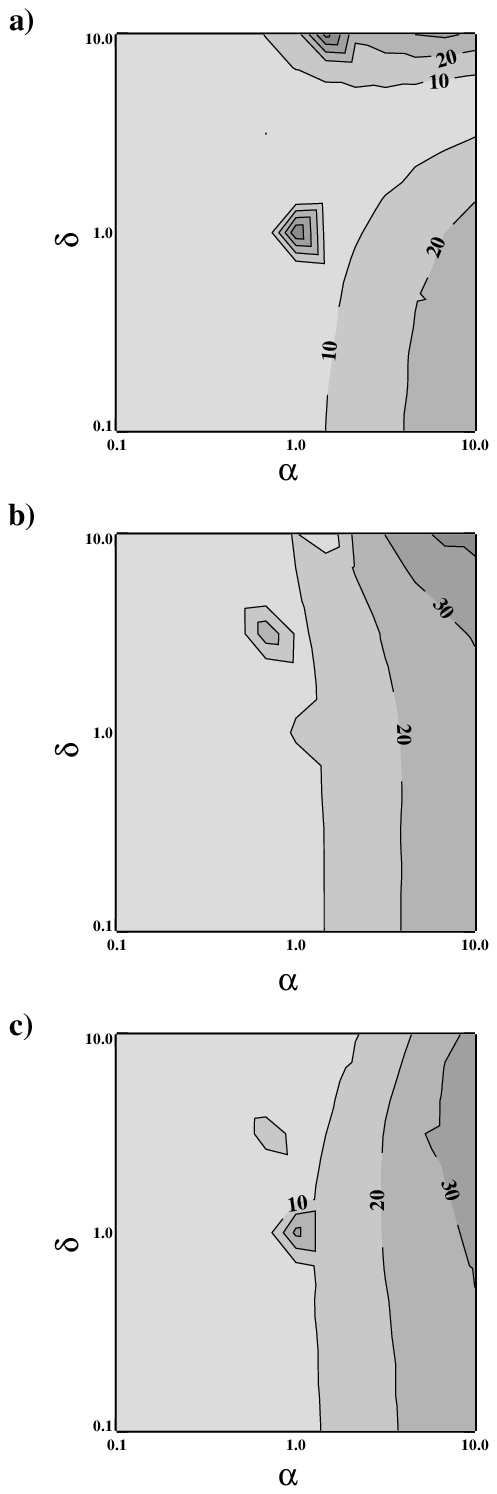}
\caption{Percentage discrepancies in results given by master equation
  approach with low truncation values for $\tau=1.0$. a: discrepancy in
  $x$(X$_2$); b:
discrepancy in $x$(XY); c: discrepancy in $x$(Y$_2$).
Darker regions indicate greater discrepancies.}
\label{lfig}\end{figure}

Following Caselli et al.\ (1998), we calculated the relative abundances of
X$_2$, XY, and Y$_2$ on the surfaces of grains as
\begin{eqnarray}
\label{xh2eq}
x({\rm H_2}) & = & \frac{\ratemolh}{\ratemolh+\rateoh+\ratemolo} \\
\label{xoheq}
x({\rm XY}) & = & \frac{\rateoh}{\ratemolh+\rateoh+\ratemolo} \\
\label{xo2eq}
x({\rm O_2}) & = & \frac{\ratemolo}{\ratemolh+\rateoh+\ratemolo}
\end{eqnarray}
We calculated the difference between the exact results, and the
 results obtained (i) through the solution of the master
equation for the low-truncation case; (ii) with the standard rate
equation approach; (iii) with the modified rate equation approach;  and
(iv) with the approximate method.  The results are plotted against
 $\alpha$ and $\delta$, for $\tau=1.0$, in Figs.\
\ref{refig} to \ref{lfig}.



Figs.\ \ref{refig} to \ref{lfig} show
the magnitudes of the percentage discrepancies between: the results
given by the standard rate equation approach, and the exact results
(Fig.\ \ref{refig}); the results given by the modified rate 
equation approach, and the exact results (Fig.\ \ref{mrefig}); the
results given by the approximate method, and the exact results (Fig.\ 
\ref{amfig}); and the results given by the master equation approach
for the low truncation case, and the exact results (Fig.\ \ref{lfig}).

Due to lack of space, the results are plotted against $\alpha$ and
$\delta$ for $\tau=1.0$ only.  Results for other values of $\tau$ can
be found at {\ttfamily{http://ast.leeds.ac.uk/$\sim$jglr}}.
In each of Figs.\ \ref{refig} to \ref{lfig}, plot (a) shows the 
discrepancies in $x$(X$_2$);  plot (b)
shows the discrepancies in $x$(XY); and plot (c) shows
the discrepancies in $x$(Y$_2$).


\begin{figure*}[!h]
\includegraphics[angle=0,width=18cm]{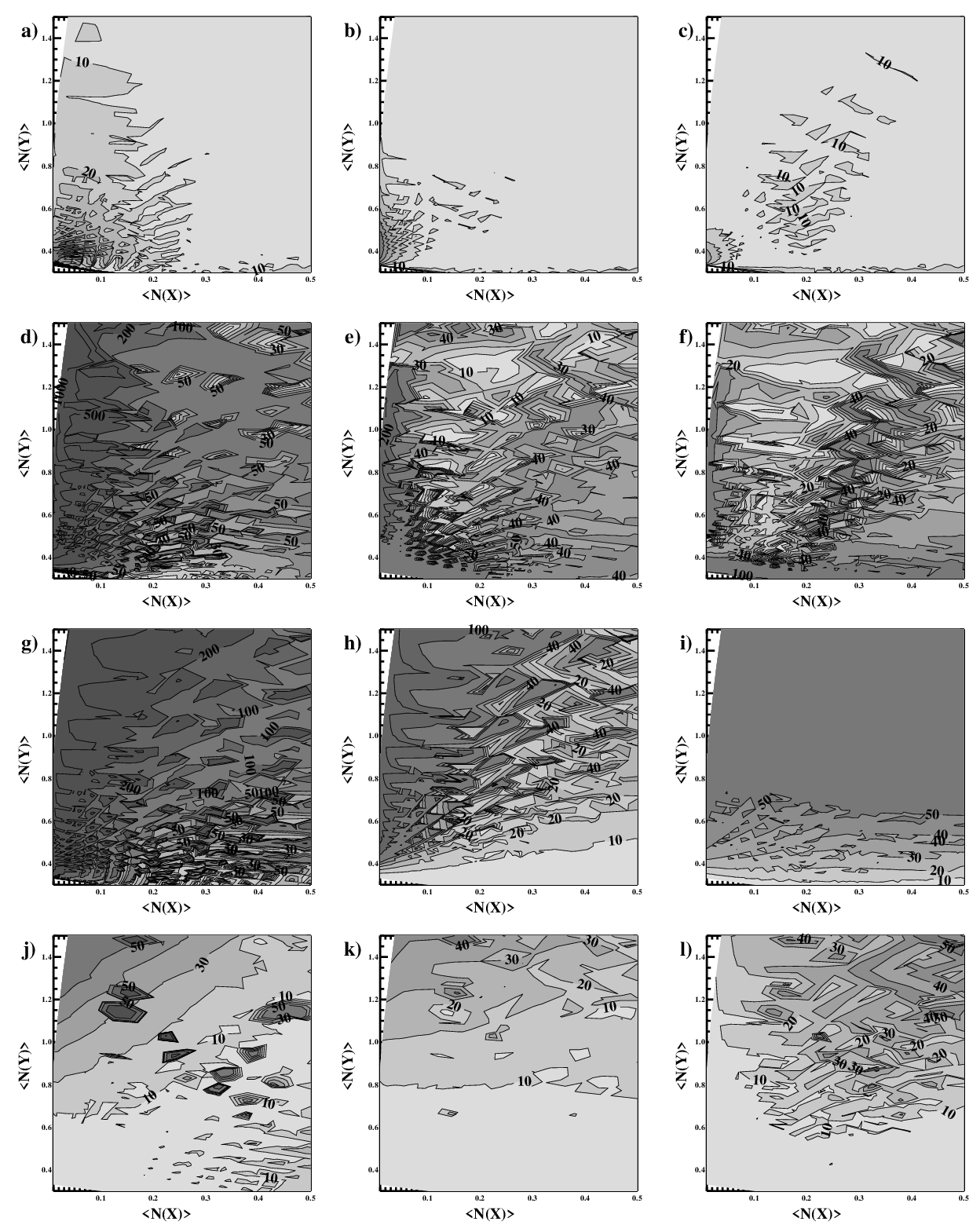}
\caption{Percentage discrepancies in results given by the various
  approaches, plotted against $\nh$ and $\no$ for all
  $\tau$.
  a -- c: standard rate equation approach;  d -- f: modified rate
  equation approach;  g -- i: approximate method;  j -- l:
  low-truncation master equation approach.  a, d, g, and j:
  discrepancy in $x$(X$_2$); b, e, h, and k: discrepancy in $x$(XY);
  c, f, i, and l: discrepancy in $x$(Y$_2$).  Darker regions indicate
  greater discrepancies.}
\label{nhnofig}
\end{figure*}

\begin{figure*}[!h]
\includegraphics[angle=0,width=18cm]{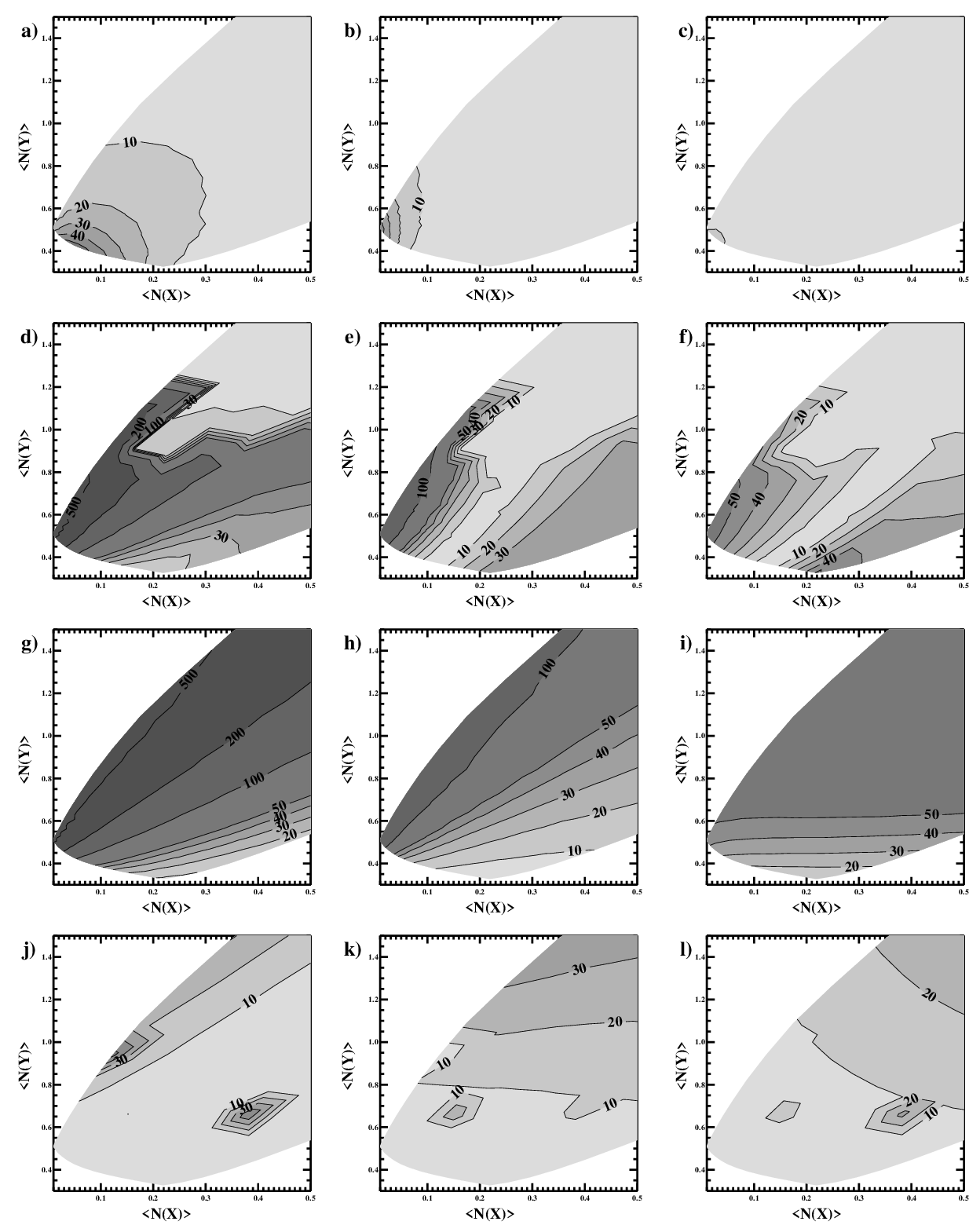}
\caption{Percentage discrepancies in results given by the various
  approaches, plotted against $\nh$ and $\no$ for $\tau=1$.
  a -- c: standard rate equation approach;  d -- f: modified rate
  equation approach;  g -- i: approximate method;  j -- l:
  low-truncation master equation approach.  a, d, g, and j:
  discrepancy in $x$(X$_2$); b, e, h, and k: discrepancy in $x$(XY);
  c, f, i, and l: discrepancy in $x$(Y$_2$).  Darker regions indicate
  greater discrepancies.}
\label{nhnot1fig}
\end{figure*}

In Fig.\ \ref{nhnofig}, the percentage discrepancies are plotted
against $\nh$ and $\no$, the average numbers of X and Y
atoms on the surface of a grain, as given by the exact method and
Eqs.\ (\ref{nheqn}) and (\ref{noeqn}).  
The white gaps at the top left-hand and bottom left-hand corners of
the plots in Fig.\ \ref{nhnofig} are caused by the fact that we did
not explore regions of $\alpha$--$\delta$--$\tau$ space that yielded
these values of $\nh$ and $\no$.  The uneven contours
in Fig.\ \ref{nhnofig} are caused by the fact that any point in
$\nh$--$\no$ space does not correspond to a single point in
$\alpha$--$\delta$--$\tau$ space, and therefore the discrepancies do
not depend only on $\nh$ and $\no$.
In Fig.\ \ref{nhnot1fig} the discrepancies are
plotted against $\nh$ and $\no$ for $\tau=1.0$ only.  Again
the white areas are regions of $\nh$--$\no$ space to which
no point in $\alpha$--$\delta$ space corresponds for $\tau=1.0$.  In
Figs.\ \ref{nhnofig} and \ref{nhnot1fig}, plots (a), (b) and
(c) are for the standard rate equation approach, plots (d), (e), and
(f) for the modified rate equation approach, plots (g), (h), and (i)
for the approximate method, and plots
(j), (k), and (l) for the low-truncation case of the master equation
approach.  Plots (a), (d), (g),
and (j) give the percentage discrepancy in $x$(X$_2$), plots (b), (e),
(h), and (k) the percentage discrepancy in $x$(XY), and plots (c),
(f), (i), and (l) the percentage discrepancy in $x$(Y$_2$).  In
Figs.\  \ref{nhnofig} and \ref{nhnot1fig}, the ranges of $\nh$--$\no$
are different because there is no desorption of Y, so X and Y are not
symmetric.

In all of Figs.\ \ref{refig} to \ref{nhnot1fig}, lighter regions
indicate lower percentage discrepancies, and darker regions higher
percentage discrepancies.

Although only the results for $\tau=1.0$ are shown in Fig.\
\ref{refig}, it was found that  the results given by the
rate equation approach are inaccurate for the case of low $\alpha$, high
$\delta$, and low $\tau$,  
corresponding to situations in which the rates of
accretion are low and the rates of X desorption, and  
of Reactions (\ref{ohprod}) and (\ref{o2prod}), are high,
so that the average surface population on a grain is low.  This
conslusion is confirmed by Figs.\ \ref{nhnofig}a--c and
\ref{nhnot1fig}a--c, which show that this approach is least accurate
for small $\nh$ and $\no$.  These are the conditions under which it is
known that the rate equation approach breaks down.

The inaccuracies in the results given by the modified rate equation
approach are shown  in Fig.\ \ref{mrefig} for $\tau=1.0$, and in Figs.\
\ref{nhnofig}d--f and \ref{nhnot1fig}d--f.  In most cases these
results are no more accurate than those of the standard rate equation
approach, and sometimes the accuracy is much less.  
Stantcheva et al.\ (2002) studied the system of H and O reacting on
the surfaces of grains to form H$_2$, OH, and O$_2$, and 
gave in their Figs.\ 1 to 3 results for the
values of $x$(H$_2$), $x$(OH), and $x$(O$_2$) calculated by various
different methods including the modified rate equation approach and
the master equation method.  The modified rate equation approach they
used was the same as that used by Stantcheva et al.\ (2001), which was
based on that used by Caselli et al.\ (1998).
They found good agreement between results
obtained with these
two approaches for the parameters they considered, which were mostly
rather different from those used by us.  
They used the same values for \accrh, \accro, \desh, \tsh, and \tso as
Hasegawa et al.\ (1992), Caselli et al.\ (1998), and Green et
al.\ (2001).  The accretion rate they used for oxygen was different
from that used for hydrogen. Their parameters were equivalent to
$10^{-12}<a_{\mathrm{H}}t_{\mathrm{s}}(\mathrm{H})<10^{-5}$;
$a_{\mathrm{O}}t_{\mathrm{s}}(\mathrm{H})\sim 10^{-10}$;
$d_{\mathrm{H}}t_{\mathrm{s}}(\mathrm{H})\sim10^{-8}$; 
and $\frac{t_{\mathrm{s}}(\mathrm{O})}{t_{\mathrm{s}}(\mathrm{H})}\sim
10^{9}$.  This corresponds to a region below and to the left of that
shown in the plots in
Fig.\ \ref{mrefig}, and for a higher value of $\tau$.
It can be seen that towards the bottom left-hand corner of Fig.\
\ref{mrefig}a, the results are more accurate than elsewhere, which is
consistent with the findings of Stantcheva et al.\ (2002).

Fig.\ \ref{amfig} shows that the approximate method gives inaccurate
results for $x$(X$_2$)  when $\delta$ is high.
If the desorption rate is high, the assumption that an X atom will
always react if it accretes onto a grain that is already populated is
invalid.  It was also found that
there are inaccuracies in the values calculated for $x$(Y$_2$) for high
$\alpha$, for all values of $\tau$.  In these cases, the accretion
rate is comparable to, or greater than, the rate of reaction of Y,
and the population of grains containing several Y atoms is non-negligible.
At high values of $\tau$, the results were found to be
inaccurate for a wide range
of values of $\alpha$ and $\delta$, particularly for $x$(X$_2$) and
$x$(Y$_2$); $x$(XY) is only inaccurate for high $\delta$ and low
$\alpha$.  The approximate
equations were constructed under the assumption that if there are two
atoms of species Y on the surface of a grain, and a third accretes,
Y$_2$ will not be formed.  If we instead assume that a grain surface
containing no atoms of X and more than two atoms of Y will always be a
site of Y$_2$ formation, Eq.\ (\ref{amo2}) should be replaced with
\begin{equation}
\ratemolo
\approx\frac{\koo+\alpha}{\koo+2\alpha}
\frac{\alpha}{2} P (0,0) 
\end{equation}
In the case of $\alpha=0.3$, $\delta=0.1$, $\tau=10$, this results in
an improvement of a factor of about 2.4 in the value obtained for the
production rate of Y$_2$.
This results in X$_2$, XY, and Y$_2$ abundances, calculated with
Eqs.\ (\ref{xh2eq}) to (\ref{xo2eq}), which differ from the exact results
by 42\%, 3\%, and 26\%, respectively;  for X$_2$ and Y$_2$ this is a
vast improvement from the results obtained when $\ratemolo$ was
calculated with Eq.\ (\ref{amo2}).
The inaccuracy at
high values of $\tau$ also causes an inaccuracy in $x$(Y$_2$) at high
$\no$, which can be seen in Fig.\ \ref{nhnofig}i, and probably
contributes to the inaccuracies seen in $x$(X$_2$) and $x$(XY) in
Figs.\ \ref{nhnofig}g and \ref{nhnofig}h, at least at
higher values of $\no$.
In Fig.\ \ref{nhnofig}g--i, one can see that the results given by this
approach for $x$(X$_2$) are inaccurate for low $\nh$, at all
values of $<N$(Y)$>$; at higher values of $\no$ the results are
inaccurate for a larger range of $\nh$.  The results for $x$(XY)
are also inaccurate for low $\nh$ for a range of $\no$.

Fig.\ \ref{lfig} shows that the inaccuracy in the results given by the
master equation method with low truncation values is high for high
$\alpha$.  In particular, the inaccuracy was found to be high in the
case of high
$\alpha$, low $\delta$, and low $\tau$.  This corresponds to the case
of a high rate of accretion, a low rate of hydrogen desorption, and 
high rates for Reactions (\ref{ohprod}) and (\ref{o2prod}).  This is the
case in which there is a large population of X atoms on the surfaces
of the grains. The inaccuracy is also high for high $\alpha$, high
$\delta$, and high $\tau$, which corresponds to fast rates of
accretion and hydrogen desorption, and a slow rate for Reaction
(\ref{o2prod}).  This results in a large population
of Y atoms on the surfaces of the grains.  The results given by the
master equation method with low truncation values are therefore
inaccurate when there are large numbers of atoms on the surfaces of
grains, as should be expected;  this can also be seen in Figs.\
\ref{nhnofig}j--l and \ref{nhnot1fig}j--l.

\begin{figure}[!h]
\includegraphics[angle=0,width=7cm]{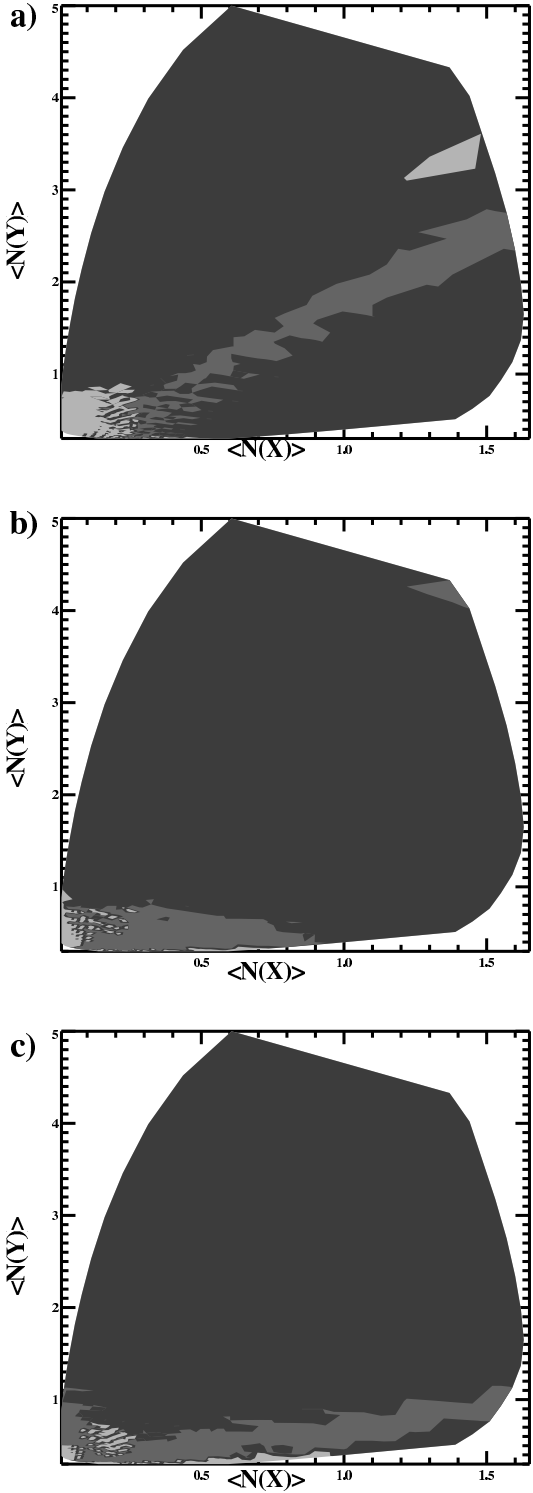}
\caption{Regions where results for (a) X$_2$, (b) XY, and (c) Y$_2$,
  given by standard rate equation approach (darker regions)
  and low-truncation master equation approach (lighter regions) are
  more accurate.  Regions of intermediate shading are where both
  approaches give results which are accurate to within 10\%.}
\label{accmethfig}
\end{figure}

Fig.\ \ref{accmethfig} shows the regions of $\nh$--$\no$ space
in which the standard rate equation approach, or the low-truncation
master equation approach, is more accurate.  Plot (a) is for X$_2$,
plot (b) for XY, and plot (c) for Y$_2$.   Dark regions indicate that
the rate equation approach is more accurate; light regions are those
in which the low-truncation master equation approach is more reliable.
Regions of intermediate
shading indicate that both approaches give results which are accurate
to within 10\%.  As expected, when either the results given by one
method or both methods are not accurate to within 10\%, the
low-truncation master equation approach is more accurate at low values
of $\nh$ and $\no$, and the rate equation approch at higher
values.  The better of the two approaches gives results for
$x$(X$_2$), $x$(XY), and $x$(Y$_2$) which are usually within 10 or
20\% -- and always within 25\% -- of those given by the exact method, 
except for the results given for $x$(Y$_2$)
in a small region of $\alpha$--$\delta$--$\tau$ space around
$\alpha=0.7$, $0.1<\delta<0.3$, $\tau>0.7$.  In these cases, the
more accurate method is the rate equation approach, and the value of
$x$(Y$_2$) given by that method is around 40\% less than the exact value.  

Although in general the rate equation approach is more accurate than
the low-truncation master equation approach for higher values of
$\nh$ and $\no$, it can be seen in Fig.\ \ref{accmethfig} that
for $\nh\sim1.2-1.4$, $\no\sim3$, the low-truncation master
equation approach is more accurate for X$_2$.   However, in most of
these cases the results given by the two methods are very similar, and
the discrepancies between them and the exact results are within about 20\%.

\section{Conclusions}
\label{conc}

We have studied grain surface chemistry with a variety of approaches
with a range of parameters.  We deliberatelty made our study as general
as was reasonable given the amount of work involved in a
multiparameter study.  We did so in anticipation of revisions of rates
pertaining to specific problems treated in earlier literature, and of
future studies of surface astrochemistry in a very wide range of
environments including some in which atomic hydrogen accretion is unimportant.

The modified rate equation method was found to be no more
accurate than the standard rate equation approach under the conditions
examined;  indeed it was in many cases less accurate.  The modified
rate equation method that we used is closely related to that of
Caselli et al.\ (2002).  They studied a system in which reaction
barriers are important, and the values of $\alpha$ approriate for
their studies are large.  The modified rate equation approach that we
used gives reliable results for large $\alpha$ for some values of
$\tau$. Hence, our findings are in harmony with those of Caselli et
al.\ (2002). The modified rate equation approach that we adopted
differs from that employed by Stantcheva et al.\ (2001) which they
found to give reliable results for the specific $\alpha\ll 0.1$ and
large $\delta$ cases they examined. They, like the authors of a number
of papers referenced here, were examining a region of parameter space
they thought to be relevant when hydrogen accretion is important for
the surface chemistry and the surface reactions proceed without
barriers. Of the approximate methods we have examined, only the
low-order truncated master equation approach is reliable for general
small $\alpha$, large $\delta$ cases.

The approximate method of Green et al.\ (2001), which was devised
specifically for the H,O system, was also found to give inaccurate
results under certain conditions, but can be improved by considering
the possibility that an atom of species Y may accrete onto the surface
of a grain on which there are already two Y atoms.

The standard rate equation approach was found to work well except
under conditions which lead to there being a small average number of X
and Y atoms on the surface of a grain.  Under these conditions, the
master equation approach with low truncation values is 
accurate.  Therefore, a combination of the two approaches should be
enough for a reasonably accurate calculation of the X$_2$, XY, and Y$_2$
abundances in the parameter regime we considered -- generally within
about 20\% of the exact values.  Practical approaches to the use of
stochastic models are under investigation (Stantcheva et al.\ 2001).

\begin{acknowledgements}
JGLR was supported by grants from the Leverhulme Trust and PPARC. TT was
supported by a grant from PPARC.

\end{acknowledgements}

{}

\begin{thebibliography}{}

\bibitem{ar77} Allen M., \& Robinson G.W., 1977, ApJ, 212, 396

\bibitem{b01} Biham O., Furman I., Pirronello V., \& Vidali G., 2001,
  ApJ, 553, 595

\bibitem{caselli98} Caselli P., Hasegawa T.I., \& Herbst E., 1998, ApJ 495, 309
\bibitem{caselli02} Caselli P., Stantcheva T., Shalabiea O., Shematovich V.I.,
 \& Herbst E., 2002, Planetary \& Space Science 50, 1257
\bibitem{c98} Charnley S.B., 1998, ApJ 509, L121
\bibitem{c01} Charnley S.B., 2001, ApJ, 562, L99 
\bibitem{ctr97} Charnley S.B., Tielens A.G.G.M., \& Rodgers S.D., 1997, ApJ 482, L203
\bibitem{green01} Green N.J.B. et al., 2001, A\&A, 375, 1111
\bibitem{h92} Hasegawa T.I., Herbst E., \& Leung C.M., 1992, ApJS 82, 167
\bibitem{pw77} Pickles J.B., \& Williams D.A., 1977, Ap\&SS, 52, 433
\bibitem{sch98} Shalabiea O.M., Caselli P., \& Herbst E., 1998, ApJ 502, 652
\bibitem{sch01} Stantcheva T., Caselli P., \& Herbst E., 2001, A\&A 375, 673
\bibitem{ssh02} Stantcheva T., Shematovich V.I., \& Herbst E., 2002,
  A\&A 391, 1069 
\bibitem{ta87} Tielens A.G.G.M., \& Allamandola L.J., 1987, In: Hollenbach
  D.J., \& Thronson H.A., (eds.) Interstellar Processes, Reidel,
  Dordrecht, p.397
\bibitem{tc97} Tielens A.G.G.M., \& Charnley S.B., 1997, Origins Life Evol. 
\bibitem{th82} Tielens A.G.G.M., \& Hagen W., 1982, A\&A, 114, 245

\end{thebibliography}
\end{document}